\documentclass[12pt]{article} 
\usepackage{amssymb} 
\usepackage{amsmath}
\usepackage{amsfonts}

\newcommand{\be}{\begin{equation}}
\newcommand{\bea}{\begin{eqnarray}}
\newcommand{\eea}{\end{eqnarray}}
\newcommand{\nn}{\nonumber}
\newcommand{\kt}{\rangle}

\newcommand{\ed}{\end{document}}

\begin{document}

\title{On the Dynamical Invariants and the Geometric Phases for a General Spin System in a Changing Magnetic Field}
\author{Ali Mostafazadeh\thanks{E-mail address: 
amostafazadeh@ku.edu.tr}\\ \\
Department of Mathematics, Ko\c{c} University,\\
Rumelifeneri Yolu, 80910 Sariyer, Istanbul, Turkey}
\date{ }
\maketitle

\begin{abstract}
We consider a class of general spin Hamiltonians of the form $H_s(t)=H_0(t)+H'(t)$ where $H_0(t)$ and $H'(t)$ describe the 
dipole interaction of the spins with an arbitrary time-dependent magnetic field and the internal interaction of the spins, 
respectively. We show that if $H'(t)$ is rotationally invariant, then $H_s(t)$ admits the same dynamical invariant as $H_0(t)$. 
A direct application of this observation is a straightforward rederivation of the results of Yan et al 
[Phys.\ Lett.\ A {\bf 251}(1999) 289 and {\bf 259} (1999) 207] on the Heisenberg spin system in a changing magnetic
field.
\end{abstract}

\baselineskip=24pt

\section{Introduction}

In Ref.~\cite{yan}, the authors construct a dynamical invariant \cite{lewis-riesenfeld} for the Heisenberg spin system in 
a changing magnetic field.\footnote{See also \cite{yan-1}.} This invariant involves two auxiliary functions that satisfy a system of 
coupled first order differential equations. These same differential equations arise in the construction of a 
dynamical invariant for the dipole interaction of a single spin in a changing magnetic field. This observation together with 
the more recent results of Ref.~\cite{p36} on the characterization of the quantum systems admitting the same dynamical 
invariant are the main motivation for the present study. 

The Hamiltonian $H_{\rm Heisenberg}$ for the Heisenberg spin system in a changing magnetic field $\vec B(t)$ is a special 
case of the spin Hamiltonians of the form
	\be
	H_s(t)=H_0(t)+ H'(t)\,,
	\label{H=H+H}
	\end{equation}
where $H_0(t)$ is the dipole interaction Hamiltonian given by
	\be	
	H_0(t):=\vec B(t)\cdot\sum_{i=1}^N\vec S_i=\sum_{i=1}^N\sum_{\alpha=1}^3B^\alpha(t)S^\alpha_i\,,
	\label{H0}
	\end{equation}
$\vec S_i=(S_i^1,S_i^2,S_i^3)$ is the spin operator for the $i$-th particle, $N$ is the number of particles, and $H'(t)$ 
is the Hamiltonian corresponding to the internal spin interaction of the system. For $H_{\rm Heisenberg}(t)$, $H'(t)$ is a 
time-independent Hamiltonian given by
	\bea
	H'_{\rm Heisenberg}&=&-A H_1,~~~~H_1:=
	\sum_{i_1,i_2}\sum_{\alpha_1,\alpha_2} Q^{\alpha_1\alpha_2}_{i_1,i_2} S_{i_1}^{\alpha_1} S_{i_2}^{\alpha_2},
	\label{h-prime-heisenberg}\\ 
	Q_{i_1i_2}^{\alpha_1\alpha_2}&:=&\left\{\begin{array}{cc}
	\delta_{\alpha_1,\alpha_2}&\mbox{if $i_1$ and $i_2$ label particles that are nearest neighbors}\\
	0&\mbox{otherwise,}\end{array}\right.
	\label{Q}
	\eea
where $A$ is a constant and $\delta_{a,b}$ denotes the Kronecker delta function. 

The purpose of this article is to show that any Hamiltonian of the form~(\ref{H=H+H}) admits a dynamical invariant that is also a 
dynamical invariant of the dipole Hamiltonian~(\ref{H0}) provided that $H'(t)$ has rotational invariance. 

First, we recall that by definition a dynamical invariant $I(t)$ for a Hamiltonian $H(t)$ is a (nontrivial) solution of the 
Liouville-von-Neumann equation:
	\be
	\frac{d}{dt}\,I(t)=i[I(t),H(t)]\;,
	\label{l-v-n}
	\end{equation}
and that any dynamical invariant satisfies
	\be
	I(t)=U(t)I(0)U^\dagger(t)\;,
	\label{q0}
	\end{equation}
where $U(t)={\cal T}e^{-i\int_0^t H(t')dt'}$ is the evolution operator generated by the Hamiltonian~$H(t)$ and ${\cal T}$
is the time-ordering operator.

It is well-known \cite{spin,nova} that one can construct a dynamical invariant $I_i(t)$ for a spin in a changing magnetic 
field using the ansatz $I_i(t)=\vec{ R}(t)\cdot \vec S_i$. Substituting this expression for $I(t)$ in Eq.~(\ref{l-v-n}) and 
using the relation $H(t)=\vec B(t)\cdot\vec S_i$ for the Hamiltonian and the $su(2)$ algebra: 
$[S_i^\alpha,S_i^\beta]=i\sum_{\gamma=1}^3\epsilon_{\alpha\beta\gamma}S_i^\gamma$, one obtains a system of first order 
differential equations for the functions $R^\alpha(t)$, \cite{spin,nova}. This invariant can also be expressed in the form 
$I_i(t)=W_i(t)S^3W_i^\dagger(t)$ where $W_i(t)$ belongs to the (spin representation of the) group $SU(2)$ generated by 
$S_i^\alpha$ and satisfies $W_i(0)=1$. It is also not difficult to see that in view of Eq.~(\ref{q0}), the evolution operator 
$U_i(t)$ generated by $H(t)$ satisfies $U_i(t)=W_i(t)Z_i(t)$ where $Z_i(t)$ is a unitary operator 
commuting with $S_i^3$, \cite{p36}. These results can be directly employed for the Hamiltonian $H_0(t)$ of Eq.~(\ref{H0}). 
Specifically, this Hamiltonian admits the invariant 
	\be
	I(t)=\sum_{i=1}^NI_i(t)=\vec{ R}(t)\cdot\sum_{i=1}^N\vec S_i=W_0(t)\left(\sum_{i=1}^NS^3_i\right)
	W_0^\dagger(t)=U_0(t)\left(\sum_{i=1}^NS^3_i\right)U_0^\dagger(t)=U_0(t)I(0)U_0^\dagger,
	\label{e5}
	\end{equation}
where $W_0(t)=\prod_{i=1}^NW_i(t)$ and $U_0(t)=\prod_{i=1}^NU_i(t)$ and
	\be
	I(0)=\sum_{i=1}^NS^3_i.
	\label{e6}
	\end{equation}
Note that because $S^\alpha_i$ with different values of $i$ commute, $[W_i(t),W_j(t)]=[U_i(t),U_j(t)]=0$. Furthermore, 
we have $U_i(t)=e^{iM_i(t)}$, $M_i(t):=\vec\rho(t)\cdot\vec S_i$, and $U_0(t)=e^{iM(t)}$ where $\vec\rho(t)$ are determined in 
terms of $\vec{ R}(t)$ and $M(t):=\vec\rho(t)\cdot\sum_{i=1}^N\vec S_i.$

Now, consider a spin Hamiltonian of the form~(\ref{H=H+H}) and suppose that $H'(t)$ is rotationally invariant, i.e., for all 
$\alpha\in\{1,2,3\}$,
	\be
	[H'(t),\sum_{i=1}^NS_i^\alpha]=0.
	\label{e6.1}
	\end{equation}
Then
	\bea
	[ H_0(t),H'(t)] &=&0\;,
	\label{q1}\\
	\left[ I(t), H'(t)\right] &=& 0\;.
	\label{q2}
	\eea
In view of Eqs.~(\ref{H=H+H}) and (\ref{q2}) and the fact that $I(t)$ is a dynamical invariant for $H_0(t)$, we have
	\[\frac{d}{dt}\,I(t)=i[I(t),H_0(t)]=i[I(t),H_0(t)+H'(t)]=i[I(t),H_s(t)]\;.\]
Hence $I(t)$ is also a dynamical invariant for the Hamiltonian $H_s(t)$. Furthermore, according to Eqs.~(\ref{H=H+H}) and 
(\ref{q1}), the evolution operator $U_s(t)$ generated by $H_s(t)$ is the product of those generated by $H_0(t)$ and $H'(t)$,
i.e.,
	\be
	U_s(t)=U_0(t)U'(t)\;,
	\label{q3}
	\end{equation}
where $U'(t)={\cal T} e^{-i\int_0^t H'(t') dt'}.$

As we mentioned earlier, the Heisenberg spin Hamiltonian $H_{\rm Heisenberg}$ considered in Refs.~\cite{yan,yan-1} is a 
special case of $H_s(t)$. It can also be easily checked that $H'_{\rm Heisenberg}$ is rotationally invariant. Hence 
$H_{\rm Heisenberg}$ also admits the invariant $I(t)$ of Eq.~(\ref{e5}). The invariant constructed in Ref.~\cite{yan} differs
from the invariant~(\ref{e5}) by a constant term that commutes with the Hamiltonian. Therefore, this term drops from both sides
of the defining equation~(\ref{l-v-n}). The only effect of this additional term is to change the degeneracy of the eigenvalues of 
the invariant. In fact, the most general dynamical invariant for the Hamiltonian $H_s(t)$ (and in particular for 
$H_{\rm Heisenberg}$) is given by
	\[I_s(t)=U_s(t)I(0)U_s^\dagger(t)=U_0(t)U'(t)I(0){U'}^\dagger(t)U^\dagger_0(t),\]
where $I(0)$ is a constant Hermitian operator. The invariant~(\ref{e5}) corresponds to the choice (\ref{e6}) for
$I(0)$. The invariant constructed in Ref.~\cite{yan} corresponds to the choice
	\be
	I(0)=\sum_{i=1}^NS^3_i+H_1.
	\label{yan-1}
	\end{equation}
As mentioned in Ref.~\cite{yan}, the eigenvectors of the invariant corresponding to~(\ref{yan-1}) are not known. This means
that an explicit solution of the time-dependent Schr\"odinger equation using this invariant is not possible. Unlike this
invariant the eigenvectors of the invariant~(\ref{e5}) are easily calculated; they are (tensor) products of the eigenvectors 
of $I_i(t)$, i.e., $W_i(t)|n_i\kt$ where $|n_i\kt$ are the eigenvectors of $S^3_i$ with eigenvalue $n_i=\pm 1/2$. 

We wish to conclude this article with the following remarks.
\begin{itemize}
\item[1.] Because it is the dynamical invariants that determine the geometric phases \cite{jpa,nova}, the geometric phases obtained
for the Hamiltonian $H_s(t)$ coincide with those of the dipole Hamiltonian $H_0(t)$. This is the reason why the expression 
obtained in Ref.~\cite{yan} for the geometric phases of a Heisenberg spin system in a changing magnetic field is essentially the 
same as the one for the geometric phases of a single spin in the same magnetic field. 
\item[2.] We can construct more general internal interaction Hamiltonians $H'(t)$ that are rotationally invariant and thus the 
corresponding total Hamiltonian $H_s(t)$ admits the same dynamical invariant as $H_0(t)$. For example, we can set
	\bea
	H'(t)&=&\sum_n\lambda_n(t) H_n,
	\label{h-prime}\\
	H_n&:=&\sum_{i_1,\cdots,i_{2n}=1}^N\sum_{\alpha_1,\cdots,\alpha_{2n}=1}^3
	Q_{i_1\cdots i_{2n}}^{\alpha_1\cdots\alpha_{2n}} S_{i_1}^{\alpha_1}\cdots S_{i_{2n}}^{\alpha_{2n}}\,,
	\label{Hn}
	\eea
where $n$ takes positive integer values, $\lambda_n$ are real-valued functions of $t$, and 
$Q_{i_1\cdots i_{2n}}^{\alpha_1\cdots\alpha_{2n}}$ are real coupling constants satisfying certain symmetry 
conditions. In order to state these conditions, we introduce the following abbreviated notation
	\[\tilde Q_{i_j i_k}^{\mu,\nu}:=Q_{i_1\cdots i_j \cdots i_k\cdots i_{2n}}^{\alpha_1\cdots\alpha_j
	\cdots\alpha_k\cdots\alpha_{2n}}~~~{\rm with}~~~\alpha_j=\mu,~~\alpha_k=\nu.\]
Then the above mentioned conditions take the form
	\bea
	{\rm for}~i_j=i_k:&& \tilde Q_{i_j i_k}^{\mu\nu} = \tilde Q_{i_j i_k}^{\nu\mu},
	\label{c1}\\
	{\rm for}~i_j\ne i_k:&& 
	\delta_{\mu,\gamma} \left(\tilde Q_{i_j i_k}^{\beta\nu}+\tilde Q_{i_k i_j}^{\nu\beta}\right)+
	\delta_{\nu,\gamma} \left(\tilde Q_{i_j i_k}^{\mu\beta}+\tilde Q_{i_k i_j}^{\beta\mu}\right)-\nn\\
	&& 
	\delta_{\mu,\beta} \left(\tilde Q_{i_j i_k}^{\gamma\nu}+\tilde Q_{i_k i_j}^{\nu\gamma}\right)-
	\delta_{\nu,\beta} \left(\tilde Q_{i_j i_k}^{\mu\gamma}+\tilde Q_{i_k i_j}^{\gamma\mu}\right)=0.
	\label{c2}
	\eea
These relations must be satisfied for all possible values of the labels $j,k,i_j,i_k,\beta,\gamma,\mu$, and $\nu$. They follow from the
requirement that $H_1$ commutes with the total spin operators $\sum_{i=1}^N S^\alpha_i$ and the identities
	\bea
	S_i^\alpha S_i^\beta&=&\frac{1}{4}\,\delta_{\alpha,\beta}+\frac{i}{2}\sum_{\gamma=1}^3\epsilon_{\alpha\beta\gamma}S^\gamma_i,\nn\\
	S_i^\alpha S_j^\beta&=&S_j^\beta S_i^\alpha~~~{\rm for}~~~i\ne j.\nn
	\eea
It is not difficult to generalize the conditions obtained for $H_1$ to $H_n$.

Perhaps the simplest nontrivial example that fulfils conditions~(\ref{c1}) and (\ref{c2}) is $Q_{i_1 i_2}^{\alpha_1\alpha_2}$ of Eq.~(\ref{Q}).
\end{itemize}

\section*{Acknowledgment}
I would like to thank Professor Alden Mead for his most invaluable comments and suggestions.

\ed